\documentclass[11pt]{article}
\usepackage{graphicx}

\def\SDiff{S\kern-1.5pt di \kern-1pt f \kern-1.5pt f}

\begin{document}
\begin{center}
{\bf  WMAP confirming the ellipticity in BOOMERanG and COBE CMB
maps}
\end{center}

\vspace{0.2in}

\noindent V.G.Gurzadyan$^{1,2}$, P.A.R.Ade$^3$, P. de
Bernardis$^4$, C.L.Bianco$^2$, J.J.Bock$^5$, A.Boscaleri$^6$, B.
P. Crill$^7$, G. De Troia$^4$, E.Hivon$^8$, V.V.Hristov$^{7}$,
A.L.Kashin$^1$, H.Kuloghlian$^1$, A.E.Lange$^7$, S.Masi$^4$,
P.D.Mauskopf$^3$, T.Montroy$^{9}$, P. Natoli$^{10}$,
C.B.Netterfield$^{11}$, E.Pascale$^6$, F.Piacentini$^4$,
G.Polenta$^4$, J.Ruhl$^{9}$, G.Yegorian$^1$

\vspace{0.2in}

$^1$ Yerevan Physics Institute (Armenia)

$^2$ ICRA, Dipartimento di Fisica, University La Sapienza, Roma
(Italy)

$^3$ Department of Physics and Astronomy, Cardiff (UK)

$^4$ Dipartimento di Fisica, University La Sapienza, Roma (Italy)

$^5$ JPL, Pasadena (USA)

$^6$ IFAC-CNR, Firenze (Italy)

$^7$ Caltech, Pasadena (USA)

$^8$ IPAC, Pasadena (USA)

$^9$ Department of Physics, U.C. Santa Barbara (USA)

$^{10}$ Dipartimento di Fisica, Tor Vergata, Roma, (Italy)

$^{11}$ Department of Physics, University of Toronto (Canada)

\vspace{0.2in}

{\bf Abstract} - The recent study of BOOMERanG 150 GHz Cosmic
Microwave Background (CMB) radiation maps have detected
ellipticity of the temperature anisotropy spots independent on the
temperature threshold.  The effect has been found for spots up to
several degrees in size, where the biases of the ellipticity
estimator and of the noise are small. To check the effect, now we
have studied, with the same algorithm and in the same sky region,
the WMAP maps. We find ellipticity of the same average value also
in WMAP maps, despite of the different sensitivity of the two
experiments to low multipoles. Large spot elongations had been
detected also for the COBE-DMR maps. If this effect is due to
geodesic mixing and hence due to non precisely zero curvature of
the hyperbolic Universe, it can be linked to the origin of WMAP low
multipoles anomaly.

\section{Introduction}

The CMB experiments of the recent years provided a solid
background for understanding the structure and early evolution of
the Universe.  The comparison of the results of independent
experiments has led to satisfactory mutual agreement. Excellent
statistical agreement was found by comparing measurements of the
angular power spectrum of the anisotropy detected by different
experiments. Moreover, the maps of the BOOMERanG \cite{Bern1},
\cite{Bern2}, \cite{Nett2}, \cite{Ruhl}, MAXIMA and ARCHEOPS
\cite{benoit} experiments have been recently compared "pixel to
pixel" to the high quality maps from the WMAP satellite
 and excellent agreement was found \cite{Bern3}, \cite{abroe}.

In the present article we will study the WMAP map \cite{Ben03} for
the region coinciding with the BOOMERanG one, to study the
distortion found in 150 GHz Boomerang maps. The latter showed
ellipticity for hot and cold anisotropy areas (spots) about 2.2
(2.5 for several degree areas) invariant to the temperature
threshold within its certain interval \cite{ellipse}. The effect
was detected both for small areas, i.e. containing from several to
10 pixels, but also for larger ones, with over 100 pixels. The
latter spots are larger than the scale of the horizon at the last
scattering surface. The biases of the estimator and of the noise
are negligible for the larger areas \cite{ellipse1}. If this
ellipticity effect, first detected for the COBE-DMR data
\cite{GT}, is due to the geodesic mixing \cite{GK1}, then it might
indicate a non precisely zero curvature of the Universe, and hence
be related with the origin of the low CMB multipoles anomaly
detected by WMAP (see e.g. \cite{Aur03,Efs03,Ell03,Lum03}). The
study below of the ellipticity in the WMAP maps, on the same
region observed by BOOMERanG, acts also as an additional
comparison of the results of the two experiments.

The geometry of the excursion sets of random fields including the
the ellipticity for Gaussian fluctuations have been 
studied before, and predict average ellipticity  1.4, with decrease 
towards higher thresholds (see \cite{Adler,Bond}).  

Measuring the ellipticity in the CMB maps actually implies the
estimation of the Kolmogorov-Sinai (KS) entropy of the dynamical
process which might lead to that effect (the geodesic mixing or
whatever). This is based on the essential fact that, KS-entropy
being local (in time) characteristics of the dynamical system,
enable to determine the properties of the evolution of the system.
With the CMB we have an analogous problem: having the maps, i.e.
the parameters of the moving photon beams of various temperature
at present epoch, we aim to recover their former history.

\section{CMB maps and geodesic mixing}

The property of mixing of geodesic flows in hyperbolic spaces
applied to the freely moving CMB photons can lead to ellipticity
in the maps \cite{GK1}. In a hyperbolic space of maximally
symmetric metric the geodesic flow $f^t$ is an Anosov system
\cite{Anosov} with the following remarkable properties. The
tangential space $TM_{f^t(x)}$ can be split into exponentially
converging and deviating subspaces $E^s(f^t(x)), E^u(f^t(x))$ (at
$t \rightarrow \infty$ and  $t \rightarrow -\infty$) of the
tangential space, so that
\begin{eqnarray*}
& & TM_{f^t(x)} = E^s(f^t(x)) \oplus E^u(f^t(x)) ,\\
& & df^{\tau} E^s(f^t(x)) = E^s (f^{t+\tau}(x)), \qquad df^{\tau}
E^u(f^t(x)) = E^u (f^{t+\tau}(x)),
\end{eqnarray*}
and for all $t > 0$, one has
\begin{eqnarray*}
& &
\parallel df^{t} v \parallel \leq C e^{-\lambda t} \parallel v \parallel,
\qquad v \in E^s(f^t(x));\\
& &
\parallel df^{t} v \parallel \geq C^{-1} e^{\lambda t}
\parallel v \parallel,
\qquad v \in E^u(f^t(x)),
\end{eqnarray*}
where $C  > 0, \lambda > 0$. The following from here the
exponential decay of the time correlation function of the geodesic
flow
\begin{equation}
|b(t)|\leq |b(0)| e^{-c \chi t}, c > 0.
\end{equation}
determines the strong chaotic (mixing) properties of the bundle
\cite{LMP}. Photon beams of various temperature mix by exponential
rate which will lead to the isotropization of the CMB and to
distortion of the maps. For Friedmann-Robertson-Walker Universe
the parameter $\chi$ is the KS-entropy and is determined by a
single scale, the diameter of the Universe,  $\chi=1/a$
\cite{GK1}. A different origin of the elongation in CMB maps is
certainly not excluded, and for more accurate maps the Kolmogorov
complexity of the anisotropy areas can be the most informative
descriptor for distinguishing the effect  \cite{G}.

\begin{figure}[htp]

\begin{center}

\includegraphics[width=8cm]{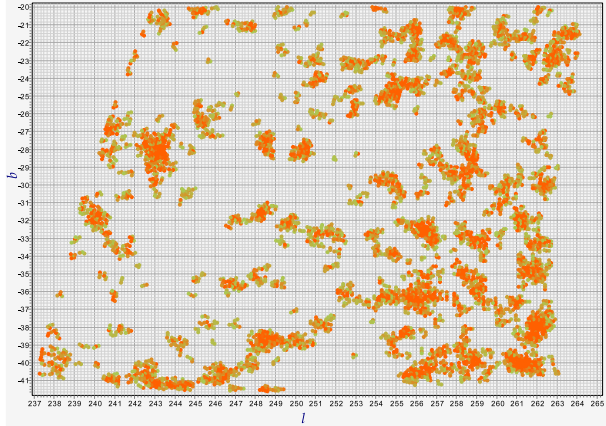}

\caption{Pixels higher than 100 $\mu K$ in the BOOMERanG "sum" or
"A+B" map obtained from three independent measurement channels at
150 GHz: A+B= B150A+(B150A1+B150A2)/2. The pixel size is 6.9
arcmin (Healpix npix=512). The measurement units ($\mu K$) refer
to thermodynamic temperature fluctuations of 2.73K blackbody.
 }

\end{center}

\end{figure}

\begin{figure}[htp]

\begin{center}

\includegraphics[width=8cm]{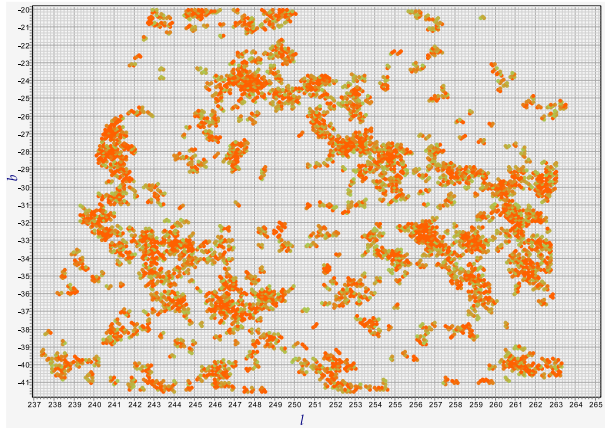}

\caption{The WMAP sum map of V and W channels for the same
thresholds as in Fig.1, and in the same sky region.
 }

\end{center}

\end{figure}

\section{Maps}

The region observed by BOOMERanG is at high Galactic latitudes,
and covers about $4 \%$ of the southern sky, with coordinates $RA
> 70^{\circ}$, $-55^{\circ} < dec < -35^{\circ}$ and $b < -20^o$.
The map analyzed here contains 33111 pixels, each of ~7 arcmin
\cite{Bern1},\cite{Bern2}, \cite{Pia}, \cite{Masi3}, \cite{Crill}.
BOOMERanG's obtained high resolution ($\sim 10'$) maps of the
microwave sky in 4 different channels, at 90, 150, 240 and 410
GHz.

Two maps from two independent detectors at 150 GHz have been used
for the ellipticity study \cite{ellipse}. Those maps have been
obtained from the time ordered data using an iterative procedure
\cite{Nato01}, which properly takes into account the system noise
and produces a maximum likelihood map. The structures of scales
larger than 10$^o$ were removed in this procedure, to avoid the
effects of instrument drifts and of 1/f noise. Three known AGN's
have been excluded from the analysis of the region.

The WMAP first-year maps have been obtained  from the publicly
available, monopole and dipole-removed datasets for channels W and
V (frequencies 61 and 94 GHz, respectively). The noise per $28'$
pixel in those maps is about 35 $\mu K$, to be compared to a noise
of $\sim 25 \mu K$ per $28'$ pixel in the 150GHz map of BOOMERanG.
For comparison of the two datasets we chose two sky regions. The
first one coincides with the BOOMERanG region; the second one has
galactic coordinates l=181$^{\circ}$.33 - 209$^{\circ}$.23 and b=
-20$^{\circ}$.03 - -41$^{\circ}$ 60. The two regions have very
similar size.

Figures 1-3 show the BOOMERANG and the two WMAP maps,
respectively, at temperature thresholds higher than $100 \mu K$.

\begin{figure}[htp]

\begin{center}

\includegraphics[width=8cm]{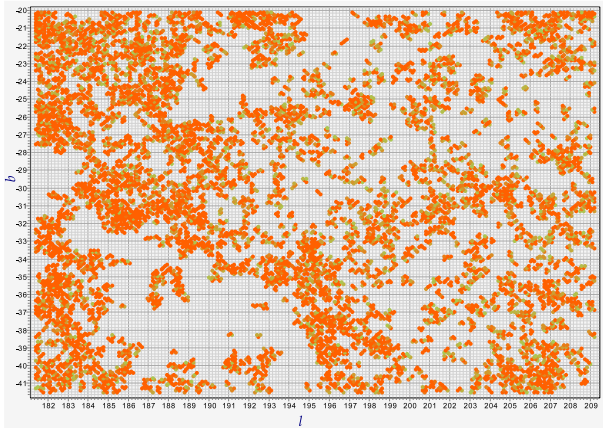}

\caption{The WMAP sum map for the same thresholds as in Fig.1, in
a sky region different from the BOOMERanG one, but with similar
size. The structure is visually different from the one of Fig.2,
due to the value of the lowest multipoles in the two regions. This
fact, however, does not affect the ellipticity results.
 }

\end{center}

\end{figure}

\begin{figure}[htp]

\begin{center}

\includegraphics[width=10cm]{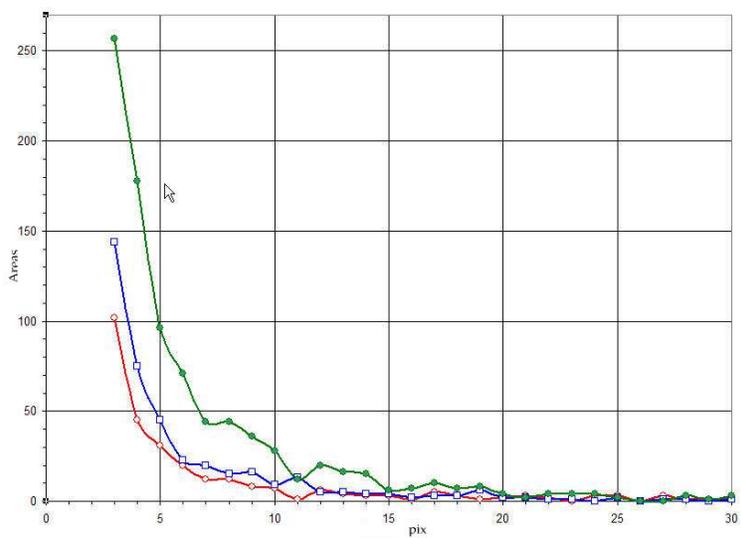}
\caption{Number of the anisotropy areas vs the size of each area
(in number of pixels) in the maps of Figs.1-3. : BOOMERanG (empty
circles), WMAP in the same region (squares), WMAP in another
region (filled circles).}

\end{center}

\end{figure}

\begin{figure}[htp]

\begin{center}

\includegraphics[width=12cm]{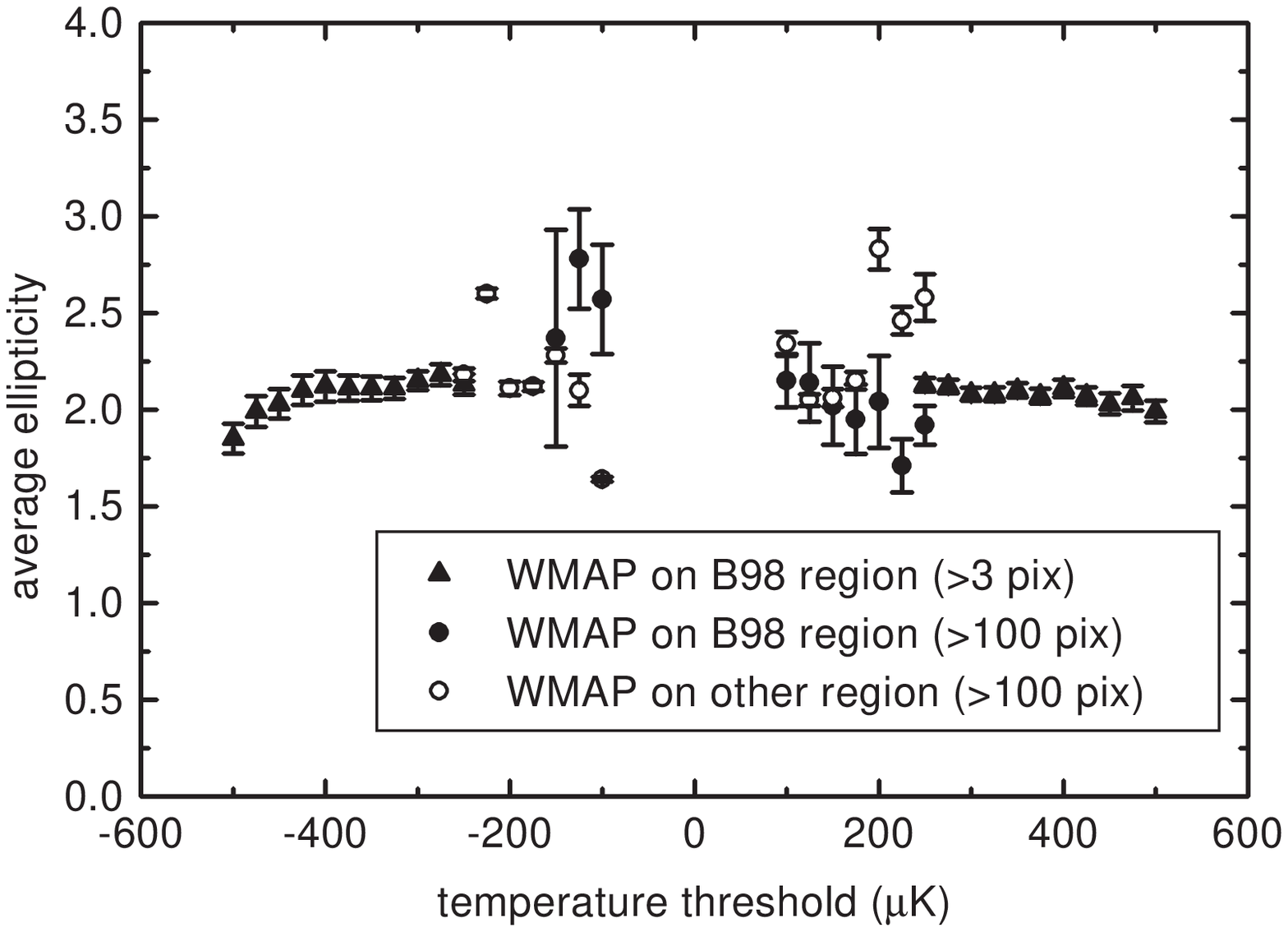}

\caption{Ellipticity vs temperature threshold  for the anisotropy
areas containing more than 3 pixels (triangles), more than 100
pixels (filled circles) in WMAP map of Fig.2, and for anisotropy
areas with more than 100 pixels in the WMAP map of Fig.3
(circles).}

\end{center}

\end{figure}

\section{Analysis}

We have used the same algorithms in the definition of the
excursion sets (areas), their centers and the ellipticities, which
were used for the BOOMERanG maps \cite{ellipse}. The study of the
same sky region will obviously reveal the differences of the
parameters of the datasets and hence their role in the studied
effect. Areas, as usual, are defined as pixel sets with
temperature equal and higher than a given temperature threshold
and lower, for negative thresholds. Fig.4 shows the number of
areas vs the size of areas (in pixels) for the 3 maps. We defined
the center as of the middle point of the coordinates of extremal
pixels of the area: we have checked that this coincides with the
center of inertia of the areas with an accuracy of few arc
minutes.

The bias of the ellipticity estimator due to the noise was up to
0.4 and up to 0.3 for BOOMERanG maps. However for larger areas
(more than 100 pixels), such bias is much smaller, less than 0.1
\cite{ellipse}. These values are applicable to the WMAP data as
well, since, as mentioned above, the noise level is comparable to
the one of BOOMERanG.

The results of the computations are shown in Fig. 5. The existence
of threshold independent ellipticity is evident, with an average
value similar to the one in the BOOMERanG maps, both for small and
large areas. The effect exists also when channel V is changed to
channel W. The obliquities of the areas did not show any preferred
direction, as it was for BOOMERanG.

Obviously the small pixel areas are clearly defined at higher
temperature thresholds and loose their identity towards lower
thresholds (forming complex connected regions). Larger areas (e.g.
more than 100-pixel) on the other hand, are not too many at higher
thresholds and hence their 'existence' threshold interval is
relatively smaller. This explains the intervals of the points on
Fig.5 (for more details see \cite{ellipse}).

While the WMAP's map in Fig.3 clearly differs from the BOOMERanG
field (Figs.1,2) due to the low multipoles, the ellipticity of the
large spots is again around 2.5, as found for BOOMERanG, with an
unavoidable scatter due to small-numbers statistics. For areas
with up to 100-pixel the WMAP data are impressive, both because of
their stability versus threshold and because of their value, fully
confirming the results of the ellipticity analysis of the
BOOMERanG map.

\section{Conclusions}

The analysis of WMAP maps in the same region observed by BOOMERanG
detects ellipticity of anisotropies of the same average value
(around 2), as  found for BOOMERanG, even though there is
difference in the maps at least due to the absence of low
multipoles in the BOOMERanG data. Ellipticity for large scale
areas had been found also in COBE-DMR maps \cite{GT}. The WMAP
data confirm the effect for scales both smaller and larger than
the horizon at the last scattering surface. This suggests that the
effect is not due to physical effects at the last scattering
surface, and can arise after, while the photons are moving freely
in the Universe. As a large-scale effect this can be related to
the WMAP low multipole anomaly, since the geodesics mixing and the
low multipoles are both related to the diameter of the Universe -
the first one via hyperbolic geometry, the second one, via
boundary conditions - and possibly even with vacuum's relevant
modes' contribution to the dark energy \cite{Lambda}.


\begin{thebibliography}{99}



\bibitem{Bern1} de Bernardis P., et al,  2000, Nature, 404, 955.

\bibitem{Bern2} de Bernardis P., et al,  2002, ApJ, 564, 559.

\bibitem{Nett2} Netterfield C.B., et al, 2002, ApJ, 571, 604.

\bibitem{Ruhl} Ruhl J., et al, 2003, ApJ, 599, 786.

\bibitem{benoit} Benoit A., et al., 2003, A\&A, 399, L19.

\bibitem{Bern3} de Bernardis P., et al, astro-ph/0311396

\bibitem{abroe} Abroe M.E., et al., 2003, astro-ph/0308355

\bibitem{Ben03} Bennett C.L., et al., 2003, ApJ Suppl. 148, 1.

\bibitem{ellipse} Gurzadyan V.G., Ade P.A.R., de Bernardis P. et al, 2003, Int.J.Mod.Phys. D, 12, 1859.

\bibitem{ellipse1} Gurzadyan V.G., Ade P.A.R., de Bernardis P. et al, astro-ph/0312305

\bibitem{GT} Gurzadyan V.G., Torres S., 1997, A \& A, 321, 19.

\bibitem{GK1} Gurzadyan V.G., Kocharyan A.A., 1992, A\&A, 260, 14; Europhys. Lett. 1993, 22, 231.

\bibitem{Aur03} Aurich R., Steiner F., 2003, astro-ph/0302264

\bibitem{Efs03} Efstathiou G., 2003, MNRAS,  343, L95.

\bibitem{Ell03} Uzan J-P., Kirchner U., Ellis G.F.R., 2003, MNRAS, 344, L65.

\bibitem{Lum03} Luminet J.-P. et al, 2003, Nature, 425, 593.

\bibitem{Adler} Adler R.J., 1981, The Geometry of Random Fields, Wiley.

\bibitem{Bond} Bond J.R., Efstathiou G., 1987, MNRAS, 226, 655.

\bibitem{Anosov} Anosov D.V. , 1967, Comm. Steklov Mathematical Inst., vol.90.

\bibitem{LMP} Lockhart C.M., Misra B., Prigogine I. 1982, Phys.Rev. D25, 921.

\bibitem{G} Gurzadyan V.G., 1999, Europhys.Lett., 46, 114.

\bibitem{Pia} Piacentini F. et al, 2002, Ap.J.Suppl., 138, 315.

\bibitem{Masi3} Masi et al 2001, ApJ 553, L93.

\bibitem{Crill} Crill B.P. et al, 2002, astro-ph/0206254

\bibitem{Nato01} Natoli P., et al., 2001, A\&A 371, 346.

\bibitem{Lambda} Gurzadyan V.G., Xue S.S. 2003, Mod.Phys.Lett. 18, 561.



\end{thebibliography}
\end{document}